\documentclass[%
 reprint,
 amsmath,amssymb,
 aps,
prl,
]{revtex4-2}
\usepackage{graphicx}
\usepackage{dcolumn}
\usepackage{bm}
\usepackage{xcolor}
\begin{document}
\preprint{APS/123-QED}

\title{Defect-induced topological interface states in rhombohedral graphite}
\author{Matteo Moioli}
\author{Marcin Mucha-Kruczynski}
 \author{Michele Pizzochero}
 \email{mp2834@bath.ac.uk}
\affiliation{Department of Physics, University of Bath, Bath BA2 7AY, United Kingdom}

\begin{abstract}
Rhombohedral graphite hosts topological surface states arising from the realization of the Su-Schrieffer-Heeger model in the out-of-plane direction. Here, we extend this picture from the surface to the bulk by envisioning stacking faults as designer defects for realizing topological interface states. Using density-functional theory and non-equilibrium Green’s function calculations, we systematically examine all symmetry-inequivalent stacking faults in rhombohedral graphite and their impact on the electronic structure. We demonstrate that these defects induce low-energy topological states localized around the fault in otherwise periodic bulk rhombohedral graphite, which exhibit a weak energy dispersion and strong sublattice polarization. We further show that electron transport is sensitive to the local stacking configuration, thus providing experimentally accessible signatures of individual stacking faults. Overall, our findings establish stacking faults as a novel route to creating topological interface states in the bulk of rhombohedral graphite via defect engineering.  \end{abstract}

\maketitle

\medskip
\paragraph{\textbf{Introduction.}} 

Rhombohedral graphite (RHG), the ABC-stacked polytype of graphite, has recently emerged as a versatile platform for the exploration of correlated and topological phenomena \cite{hagymasi2022observation,parra2026nematic,miao2026various}, including superconductivity \cite{han2025signatures,choi2025superconductivity},  the anomalous Hall effect \cite{han2024large,li2026transdimensional}, and metal-free magnetism \cite{liu2025electronic,zhou2024layer}. These phenomena originate from nearly flat bands at the Fermi energy, with the corresponding states being sublattice-polarized at the surface layers. \cite{PhysRevB.73.245426}. Their formation can be understood through the Su-Schrieffer-Heeger (SSH) model \cite{su1980soliton}. Along the out-of-plane direction, the alternating intra- and  inter-layer hopping interactions in RHG emulate a dimerized chain, whose topology is captured by a winding number. In the topologically non-trivial phase, the bulk-boundary correspondence enforces end states \cite{asboth2016short}, which in RHG manifest at the surfaces over the region of the Brillouin zone bounded by the bulk nodal ring.
Such topological states have been investigated both experimentally \cite{henck2018flat,shi2020electronic,zhang2024correlated} and theoretically \cite{slizovskiy2019films,PhysRevB.89.035431}, typically within finite-thickness RHG \cite{taut2013electronic,taut2014electronic,weht2026topological} or in RHG domains embedded in Bernal graphite \cite{sarsfield2025electronic}. Topological states residing in the bulk of RHG, in contrast, have received comparatively little attention \cite{soneji2026rhombohedral,soneji2026electron}.

In this Letter, we introduce stacking faults as designer defects for the realization of topological interface states in bulk RHG.  By combining first-principles with non-equilibrium Green's function calculations, we demonstrate that the electronic structure and quantum transport properties of these states are governed by the local stacking configuration, which acts as a knob that is not available in their surface counterparts. Our work thus establishes defect engineering as an alternative mechanism to create and probe topological states in RHG. 


\medskip
\paragraph{\textbf{Methodology.}} 

Our first-principles calculations are performed using semi-local density-functional theory (DFT), implemented in the \textsc{siesta} package \cite{soler2002siesta,garcia2020siesta} version 5.4.1.
We treat the exchange and correlation effects, together with the Van der Waals interactions, with a VdW-DF2 functional \cite{lee2010higher}. 
The Kohn-Sham wave functions of the valence electrons are represented as a linear combination of local basis functions of double-$\zeta$ plus polarization quality, in combination with a mesh cutoff of 425 Ry. Core electrons are replaced by norm-conserving pseudopotentials generated following the Troullier-Martins scheme \cite{troullier1991efficient}, obtained from the PseudoDojo library \cite{garcia2018psml}. The Fermi temperature is set to 300 K. In our models, the stacking fault comprises two or four layers, depending on the configuration, encapsulated between 24 or 21 layers of pristine RHG on either side.
Geometry optimizations are performed with a Broyden algorithm with a tolerance on forces of  0.01 eV/{\AA}. The integration over the Brillouin zone is carried out using Monkhorst-Pack grids of $k$-points of 15$\times$15$\times$1 for geometry optimizations and 60$\times$60$\times$1 for the electronic local density of states and transmission, while a 120$\times$120$\times$1 $K$-centred grid was used for energy cuts along specific directions. We have explicitly verified the convergence of our results with respect to the computational parameters.

To determine the spectral density and quantum transport of stacking faults in RHG, we combine first-principles Hamiltonians with the non-equilibrium Green's function formalism, as implemented in \textsc{transiesta} \cite{brandbyge2002density,papior2017improvements}. Post-processing and data analysis were carried out with \textsc{sisl} \cite{zerothi_sisl}. In these open-boundary calculations, the scattering region consists of 46 layers, including a central stacking fault of four layers, embedded between 21 layers of pristine RHG above and 21 below.  We obtain the $k$-resolved spectral density as $\rho_{\mathbf{k}}(E) =-\frac{1}{\pi}\operatorname{Im} G^+_{\mathrm{eff},\mathbf{k}}(E)$, where $G^{+(-)}_{\mathrm{eff},\mathbf{k}}(E)$ is the retarded (advanced) Green's function of the central scattering region. Similarly, we obtain the left-right transmission function as $T_{\mathbf{k}}^{LR}(E) =\operatorname{Tr}\left[\Gamma^L_{\mathbf{k}}(E)G^+_{\mathrm{eff},\mathbf{k}}(E)\Gamma^R_{\mathbf{k}}(E) G^-_{\mathrm{eff},\mathbf{k}}(E)\right]$, with $\Gamma^{L(R)}_{\mathbf{k}}(E)$ being the broadening matrix pertaining to the left (right) electrode $L$ ($R$), describing its coupling to the central region through the electrode self-energies. Finite-bias calculations are performed self-consistently.

\medskip
\paragraph{\textbf{Structure and stability of stacking faults.}} 

\begin{figure} [t]
\includegraphics[width=1\linewidth]{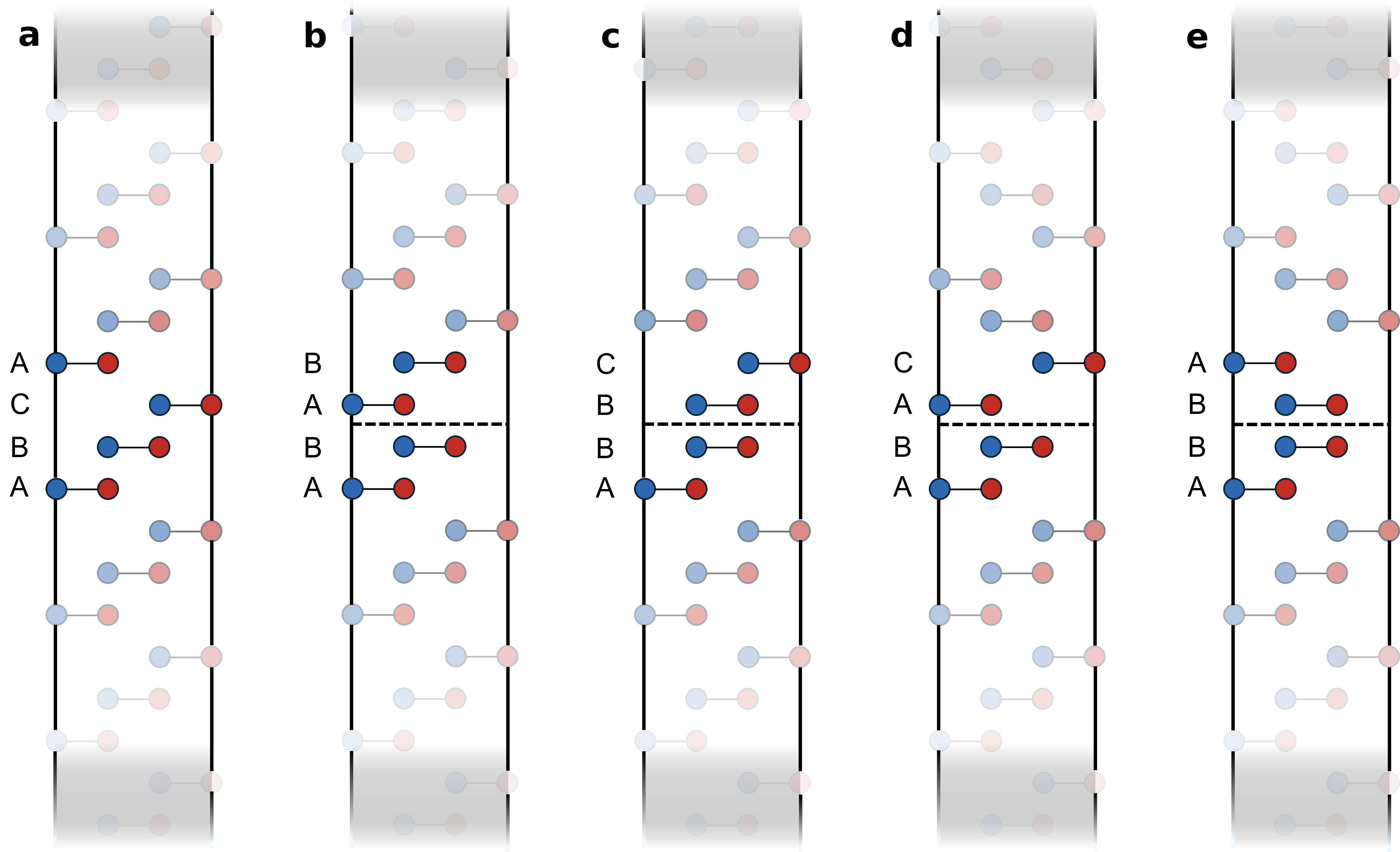}
\caption{\textbf{Atomic structure of stacking faults in RHG.} Illustration of the atomic structure of \textbf{a,} pristine RHG, denoted as $-${AB}$|${CA}$-$, as well as the  four symmetry-inequivalent stacking faults in RHG, denoted as \textbf{b,} $-${AB}$|${BC}$-$, \textbf{c,} $-${AB}$|${AB}$-$, \textbf{d,} $-${AB}$|${AC}$-$, and \textbf{e,} $-${AB}$|${BA}$-$. Blue and red circles denote carbon atoms belonging to opposite sublattices.}
\label{Figure-1}
\end{figure}

We introduce stacking fault defects in otherwise pristine RHG by considering the two  high-symmetry stacking configurations of graphite that are different from rhombohedral, that is, Bernal (i.e., AB-stacking) and  simple hexagonal (i.e., AA-stacking). These give rise to a large variety of possible defect configurations, of which only four are symmetry inequivalent \cite{soneji2026rhombohedral}, as shown in Figure \ref{Figure-1}. The resulting structures are denoted as $-${AB}$|${AB}$-$, $-${AB}$|${BC}$-$, $-${AB}$|${AC}$-$, and $-${AB}$|${BA}$-$ to highlight their different local stacking, where  hyphens denote the continuing uninterrupted rhombohedral stacking sequence. We note that each of these stacking faults can be equivalently viewed as a homojunction formed by connecting appropriately two semi-infinite crystals of RHG. 

The formation of the stacking fault introduces only a minor change to the interlayer distance in the vicinity of the defect. For stacking faults with Bernal-like stacking, $-${AB}$|${AB}$-$ and $-${AB}$|${AC}$-$, the interlayer distance at the defect is 3.35 {\AA} and 3.34 {\AA}, effectively equal to that of pristine RHG, 3.34 {\AA}. For the stacking faults with simple hexagonal-like stacking,  $-${AB}$|${BC}$-$ and $-${AB}$|${BA}$-$, the interlayer distance increases by approximately 3{\%}, to 3.43  {\AA} and 3.47  {\AA}, respectively. This is consistent with the larger interlayer distance reported for hexagonal as compared to Bernal graphite \cite{lee2008growth,birowska2011van,francas2025influence}.

To assess the thermodynamic stability of the stacking faults, we determine the change in cohesive energy upon defect formation, relative to pristine RHG. As compared to RHG, the $-${AB}$|${AB}$-$ and $-${AB}$|${AC}$-$ stacking faults are essentially equally stable, as they both lie within 0.01 meV/atom above it in energy. On the other hand, the $-${AB}$\mid${BC}$-$ and $-${AB}$\mid${BA}$-$ faults are noticeably less favorable by approximately 0.26 meV/atom each. This difference in stability reflects the interlayer separation. Stacking faults that preserve the interlayer distance spacing are energy degenerate with pristine RHG, while those that affect it incur a small yet non-vanishing energetic cost. In all cases, however, the relative cohesive energy remains below 1 meV/atom. This suggests that all of these faults could plausibly coexist in real samples.

In addition to forming as native defects, these stacking faults could be introduced deliberately, thanks to recent progress in van der Waals technology, where the mechanical manipulation of layered crystals in general \cite{Yeo2025, Lau2022, Ribeiro-Palau2018}, and graphite in particular \cite{Yang2019, Hu2022}, enables the fabrication of targeted local stacking configurations.

\medskip
\paragraph{\textbf{Electronic structure of stacking faults.}} 

\begin{figure*}[t]
\includegraphics[width=1\linewidth]{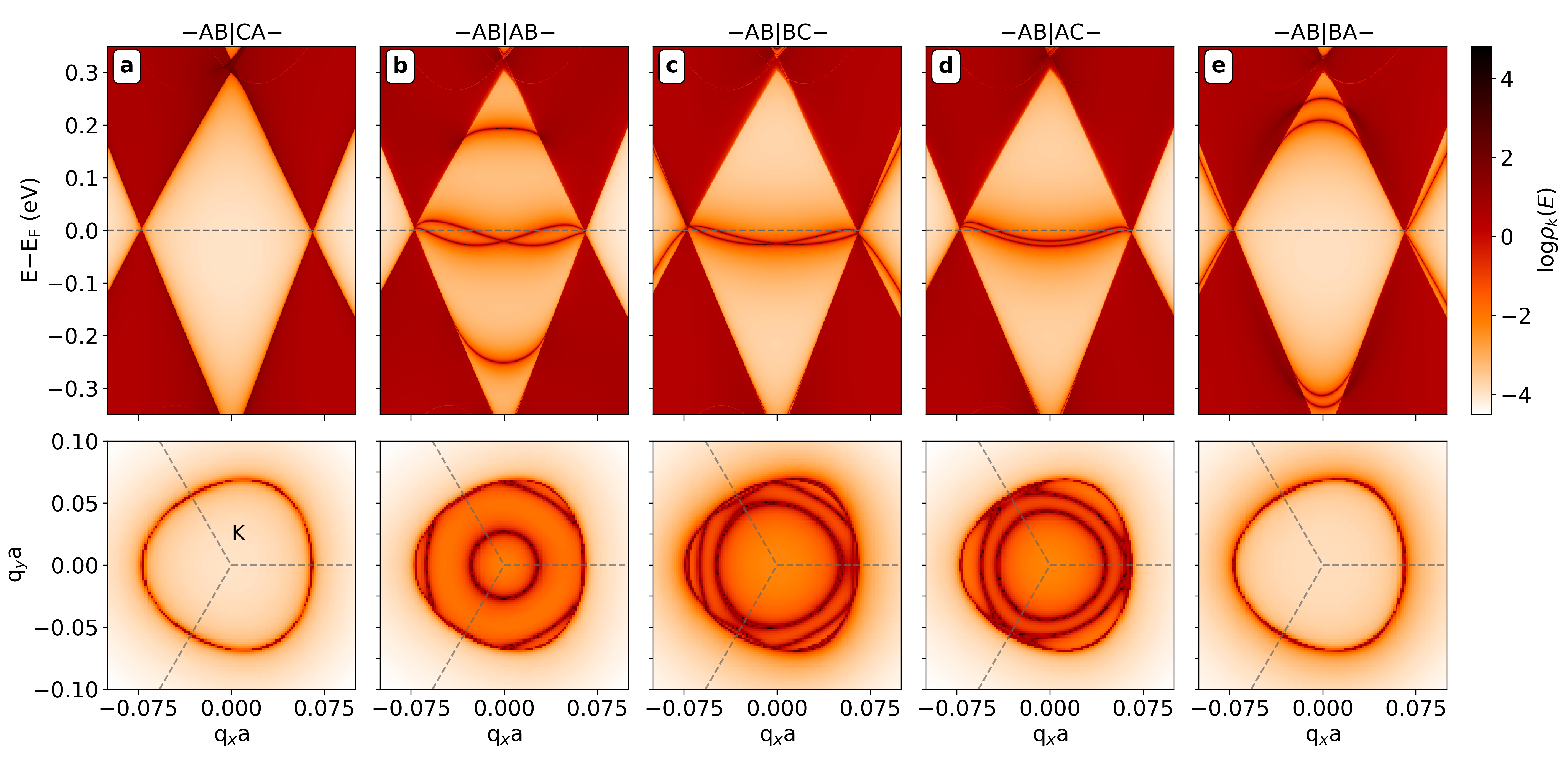}
\caption{\textbf{Electronic structure of stacking faults in RHG.} $k$-resolved local density of states as a function of $|\mathbf{q}|=|\mathbf{k}-\mathbf{K}|$ in the -0.1$<q_xa<$0.1 interval for $q_ya=0$, (top panels) and Fermi surfaces (bottom panels) for  \textbf{a,} pristine RHG, denoted as $-${AB}$|${CA}$-$, as well as the  four symmetry-inequivalent stacking faults in RHG, that is, \textbf{b,} $-${AB}$|${BC}$-$, \textbf{c,} $-${AB}$|${AB}$-$, \textbf{d,} $-${AB}$|${AC}$-$, and \textbf{e,} $-${AB}$|${BA}$-$.}
\label{Figure-2}
\end{figure*}

To determine the impact of stacking faults on the electronic structure of RHG, we obtain the $k$-resolved density of states and Fermi surface. For reference, in Figure \ref{Figure-2}(a) we show the results for pristine RHG, which exhibits Dirac-like band crossings near the corner of the Brillouin zone $K$. These crossings form a nodal ring in the $(k_x,k_y)$ plane around $K$, with threefold-symmetric warping \cite{mccann2006landau,koshino2009trigonal,kaladzhyan2021surface}. Within the SSH picture, this nodal ring marks the boundary between topologically trivial states inside the ring and topologically non-trivial states outside it, when occurring at the Fermi energy, $E_F$.

Out of the four stacking fault configurations, three of them, namely $-${AB}$|${AB}$-$, $-${AB}$|${BC}$-$, and $-${AB}$|${AC}$-$ displayed in Figures \ref{Figure-2}(b-d), present a pair of bands in the vicinity of $E_F$ with bandwidths of 47, 36, and 45 meV, respectively. This is comparable to the width of the topological surface band of RHG observed from angle-resolved photoemission spectroscopy \cite{zhang2024correlated}. For $-${AB}$|${AB}$-$ and $-${AB}$|${BC}$-$, these bands are degenerate at $K$, with their crossing occurring 21 and 26 meV below $E_F$, respectively, while for $-${AB}$|${AC}$-$ they are split in energy by 9 meV \cite{soneji2026electron}. 

The $-${AB}$|${AB}$-$ and $-${AB}$|${BA}$-$ stacking faults feature two and four interface bands located away from  $E_F$ yet still within the $k$-dependent bulk gap, respectively, as shown in  \ref{Figure-2}(b) and \ref{Figure-2}(e). In both cases, these interface bands exhibit an approximately quadratic dispersion in the vicinity of $K$. For the  $-${AB}$|${AB}$-$ stacking fault, these bands occur at energies of -252 and 193 meV relative to $E_F$ at $K$, while for the $-${AB}$|${BA}$-$ fault they occur at $-$330, $-$313, 209, and 251 meV.

\begin{figure} [htbp]
\includegraphics[width=1\linewidth]{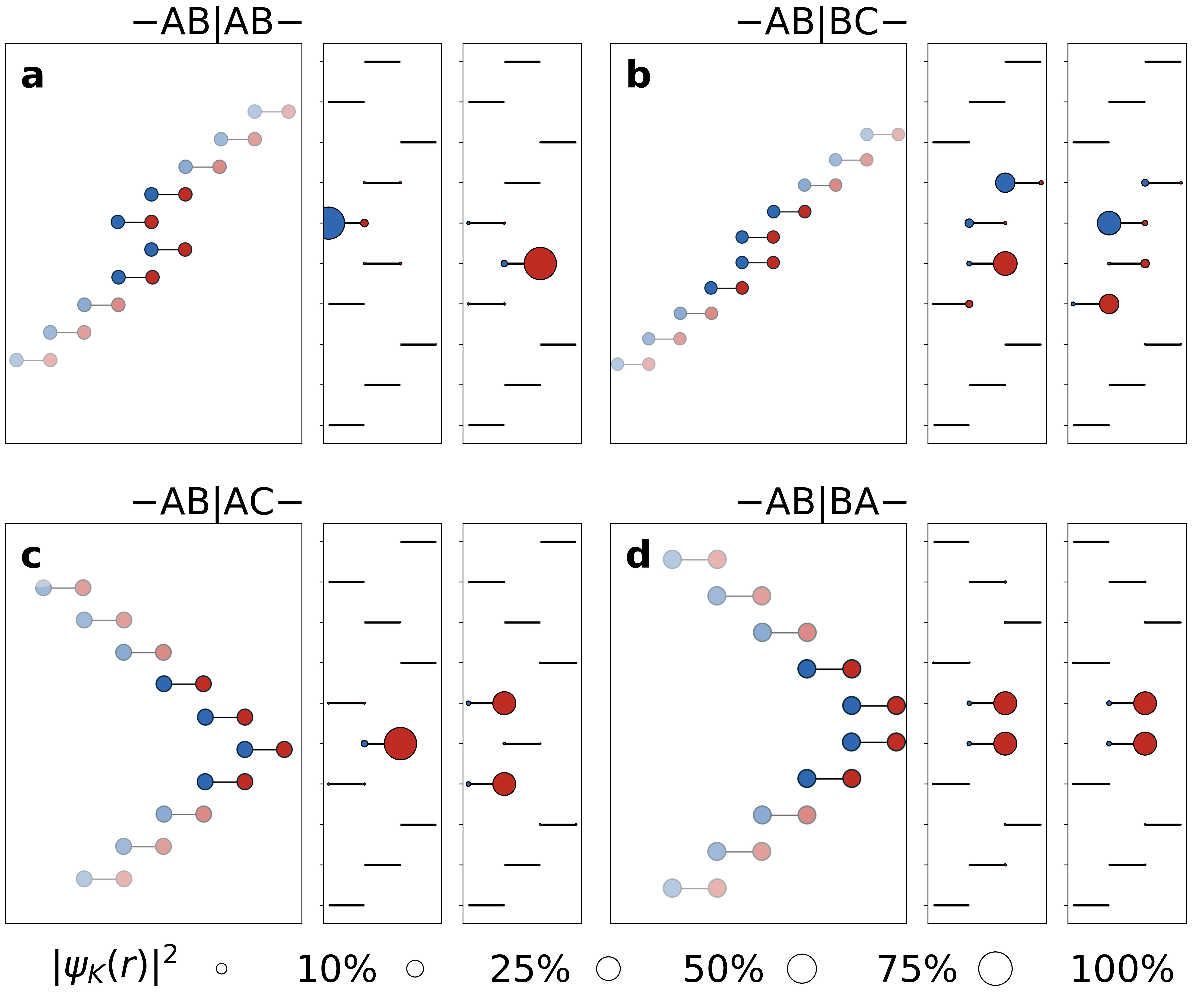}
\caption{\textbf{Localization patterns of stacking fault-induced states in RHG.}  Illustration of the atomic structure of the stacking faults in RHG, along with the layer- and sublattice-resolved  topological interface states for the \textbf{a,} $-${AB}$|${AB}$-$, \textbf{b,} $-${AB}$|${BC}$-$, \textbf{c,} $-${AB}$|${AC}$-$, and \textbf{d,} $-${AB}$|${BA}$-$  faults. }
\label{Figure-3}
\end{figure}

We gain further insight into the stacking fault-induced states by  examining their layer- and sublattice-resolved localization. This enables us to establish a direct connection between these defect states and the SSH model, supporting their topological nature.  To that end, we determine the squared modulus of the wave functions of the low-energy interface states at $K$, $|\psi_K(\mathbf{r})|^2$, and integrate their contributions over each RHG layer and sublattice. The results are reported in Figure \ref{Figure-3}, together with an illustration of  the stacking fault geometries in RHG, which readily admit an interpretation in terms of the SSH-like model \cite{soneji2026rhombohedral}. 

As shown in Figure \ref{Figure-3}(a,b), the low-energy interface states of $-${AB}$|${AB}$-$ and $-${AB}$|${BC}$-$ display a sublattice localization pattern akin to that of the end states of the SSH chain. As previously noted, a stacking fault effectively partitions the system into two coupled semi-infinite chains, each hosting an edge-localized end state. For the $-${AB}$|${AB}$-$ case, 92 and 93\% of the two low-energy states are localized on that end site. For wave vectors away from the valley centre, the edge state spreads further from the interface yet remains confined to one sublattice \cite{soneji2026rhombohedral}.
Because the edge sites of the two semi-infinite chains are not vertically aligned, their coupling proceeds via momentum-dependent hopping interactions which vanish at $K$, thereby leading to the energy degeneracy of the edge states at this wave vector. A similar localization pattern is observed for the $-${AB}$|${BC}$-$ stacking fault, except for an additional contribution arising from the other half-chain, extending across the four stacking-fault layers. Despite this more spatially extended character, we emphasize that, within each half-chain, the localisation pattern retains the characteristic SSH edge-state behavior, with the wave function residing predominantly on a single sublattice. 

\begin{figure*}[htbp]
\includegraphics[width=1\linewidth]{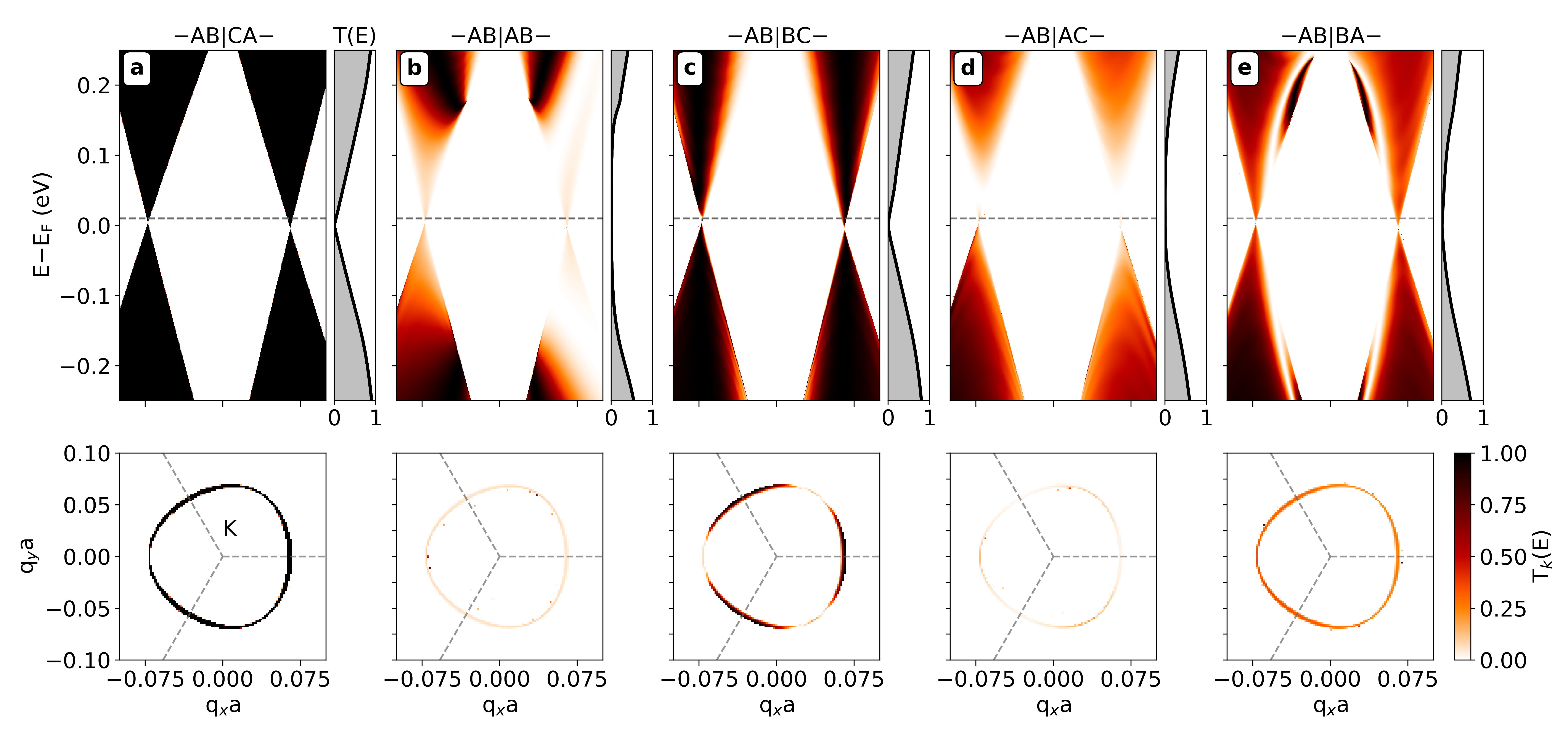}
\caption{\textbf{Zero-bias transmission function across stacking faults in RHG.} $k$-resolved, zero-bias transmission profile as a function of $|\mathbf{q}|=|\mathbf{k}-\mathbf{K}|$ in the -0.1$<q_xa<$0.1 interval for $q_ya=0$,  (top left panels), average transmission function (top right panels), and energy cut at $E-E_F=1$ meV (bottom panels) for  \textbf{a,} pristine RHG, denoted as $-${AB}$|${CA}$-$, as well as the  four symmetry-inequivalent stacking faults in RHG, that is, \textbf{b,} $-${AB}$|${BC}$-$, \textbf{c,} $-${AB}$|${AB}$-$, \textbf{d,} $-${AB}$|${AC}$-$, and \textbf{e,} $-${AB}$|${BA}$-$.}
\label{Figure-4}
\end{figure*}

In Figure \ref{Figure-3}(c, d), we show the corresponding results for the $-${AB}$|${AC}$-$ and  $-${AB}$|${BA}$-$ stacking faults. The $-${AB}$|${AC}$-$ configuration also exhibits the topological character. Due to the presence of a reflection plane passing through one of the layers, one state can be interpreted as the edge state shared by both half-chains, while the other exhibits a mirror-symmetric localization on the same sublattice across the two adjacent layers.  For the $-${AB}$|${BA}$-$ case, due to the reflection plane lying between the layers, the end atoms of both half-chains are located directly above one another. As a result, the vertical hopping interactions lift the energy degeneracy of the two states. Indeed, while the states located $-$313 and 209 meV relative to $E_F$ display a non-topological localization pattern (not shown), the states at -330 and 251 meV present the topological features of the SSH model. Despite having the same local simple-hexagonal stacking, the $-${AB}$|${BA}$-$ and $-${AB}$|${BC}$-$ stacking faults present a different low-energy physics, uniquely determined by the presence or absence of a mirror plane between the BB-stacked layers.

\medskip
\paragraph{\textbf{Quantum transport across stacking faults.}}

We next examine how stacking faults affect the quantum transport across RHG. Figure~\ref{Figure-4} shows the $k$-resolved and $k$-integrated transmission
function, along with an in-plane cut around $K$ at an energy $E - E_F=0.01$~eV, for both the pristine and defective RHG. In pristine RHG, the transmission is 
equal to one for every bulk state, such that it is proportional to the electronic density of states and increases linearly away from $E_F$, as expected.
Introducing the stacking fault depletes the transmission in the vicinity of $E_F$, yet to very different extents depending on the defect configuration. The $-$AB$|$BC$-$ fault largely preserves the pristine profile, with only a moderate reduction of the transmission along the ring. In contrast, $-$AB$|$AB$-$ and $-$AB$|$AC$-$
almost entirely suppress transmission within $|E-E_F|\lesssim0.1$~eV, with appreciable weight recovered only at higher energies. The $-$AB$|$BA$-$ fault shows an intermediate, nearly uniform reduction along the ring, along with additional high-transmission arcs and sharp minima at $|E-E_F|\gtrsim0.2$~eV, which we ascribe to the higher-energy defect states previously described. Notably, in all cases the suppression is essentially a rescaling of the pristine ring signal, in that no new transmitting channels appear at low energy and the shape of the ring is preserved.

To quantify the depletion of transmission upon defect formation, we introduce, following earlier works \cite{pizzochero2021edge,pizzochero2021quantum}, the following quantity 
$$\tau=\frac{\int_{E_1-E_F}^{E_2-E_F}T_\text{SF}(E)dE}{\int_{E_1-E_F}^{E_2-E_F}T_\text{RHG}(E)dE},$$
where $T_\text{SF}(E)$ and $T_\text{RHG}(E)$ are the $k$-averaged transmission profiles of defective and pristine RHG, respectively, integrated in energy from $E_1 - E_F$ = $-$0.1 eV to $E_2 - E_F$ = 0.1 eV. In other words, $\tau$ quantifies the fraction of transmission within 0.1 eV of $E_F$ that is retained in defective RHG.

We obtain $\tau =76\%$ for the $-$AB$|$BC$-$ fault,  $\tau =32\%$ for the $-$AB$|$BA$-$, $\tau =15\%$ for the  $-$AB$|$AC$-$, and $\tau =6\%$ for the $-$AB$|$AB$-$ fault, indicating a wide degree of transmission retention depending on the defect. Specifically, the hexagonal-like (AA-stacked) faults transmit systematically more than the Bernal-like (AB-stacked) ones, albeit being the ones inducing the most significant alteration of the interlayer distance. The transmission is therefore governed by the local stacking configuration rather than by the structural distortion. The defect states, being localized in the plane of the fault, do not themselves provide an out-of-plane conduction channel.

\begin{figure}[]
\includegraphics[width=1\linewidth]{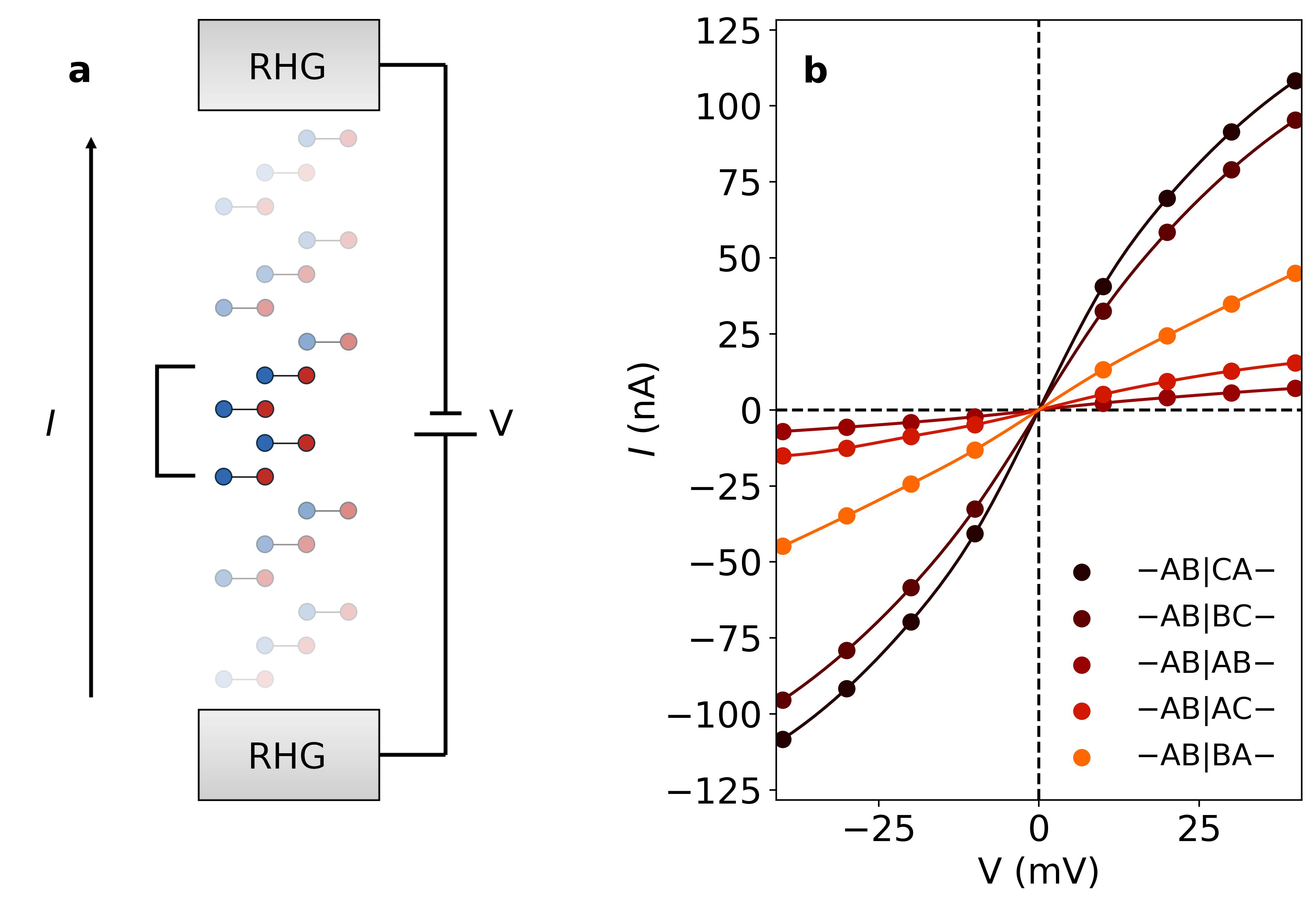}
\caption{\textbf{Finite-bias quantum transport across stacking faults in RHG.} \textbf{a,} Illustration of the device setup used in our simulations, in which a bias voltage ($V$) is applied between top and bottom electrodes to determine the resulting out-of-plane current $I$ in pristine and defective RHG. \textbf{b,} $I$-$V$ characteristics of pristine ($-${AB}$|${CA}$-$) and defective RHG for four stacking fault configurations.}
\label{Figure-5}
\end{figure}

Finally, we investigate the out-of-plane current-voltage ($I$-$V$) characteristics for $|V|\leq40$~mV, in the setup illustrated in Figure \ref{Figure-5}(a).  As shown in  \ref{Figure-5}(b), pristine RHG carries the largest current, reaching $108$~nA at $V=40$~mV, with a sublinear dependence on $V$. The stacking faults reduce the current following the same hierarchy as the equilibrium transmission: $-$AB$|$BC$-$ ($I = 95$~nA) remains close to the pristine system, followed by $-$AB$|$BA$-$ ($I = 45$~nA), $-$AB$|$AC$-$ ($I = 15$~nA), and $-$AB$|$AB$-$ ($I = 7$~nA). At $V=10$~mV, the currents amount to approximately $80\%$, $33\%$, $12\%$, and $5\%$ of the pristine value, respectively, not dissimilar from the values of $\tau$ extracted from the zero-bias transmission. The current is therefore controlled by the local stacking configuration, with the hexagonal-like faults sustaining substantially larger currents than the Bernal-like ones. Our transport picture is phase-coherent, which is justified by the out-of-plane coherence length of 2--5~nm in RHG \cite{li2026transdimensional}, exceeding the thickness of the region hosting a single fault.

The topological states introduced by stacking faults are, by construction, buried in the bulk. Unlike surface topological states in RHG, they are therefore largely invisible to experimental techniques such as angle-resolved photoemission spectroscopy and scanning tunneling. Out-of-plane transport, in contrast, is intrinsically a bulk probe, as the current must
traverse every interface along the stacking direction, thus providing a natural approach to detecting and characterising such faults, given that different stacking fault configurations yield different $I$-$V$ characteristics. In practice, a measured (differential) conductance  can therefore be compared against these fingerprints to identify the dominant class of fault, and a systematic reduction of the current relative to a defect-free reference would signal the presence of faults and constrain their density. We anticipate that this could be tested in nanoscale RHG devices, complementing structural characterisation techniques that cannot easily distinguish between stacking faults of similar energy.

\medskip
\paragraph{\textbf{Summary and conclusions.}}
In summary, we have introduced stacking faults as a novel strategy for the realization of topological states of rhombohedral graphite, extending the SSH physics known at its surfaces to the bulk. Through first-principles calculations, we showed that single faults are energetically inexpensive and host weakly dispersive, sublattice-polarized states located in the topologically non-trivial region around $K$. Their wave functions emulate the end states of two SSH half-chains, remaining at the Fermi level in most defect configurations. These defects suppress the low-energy transmission by an amount set by the local stacking, so that out-of-plane transport, unlike surface probes, may offer a route to detect and classify these faults.  

To conclude, although defects are often viewed as detrimental to the electronic properties of crystals, our findings show that stacking faults turn the bulk of rhombohedral graphite into a platform for topological interface states, with both their electronic structure and their transport signature dictated by the local stacking. This opens a route to control the topology of layered crystals via defect engineering.

\bibliography{ref_article}

@Article{Yeo2025,
author={Yeo, Youngki
and Sharaby, Yoav
and Roy, Nirmal
and Raab, Noam
and Watanabe, Kenji
and Taniguchi, Takashi
and Ben Shalom, Moshe},
title={Polytype switching by super-lubricant van der {W}aals cavity arrays},
journal={Nature},
year={2025},
month={Feb},
day={01},
volume={638},
number={8050},
pages={389-393},
}

@Article{Lau2022,
author={Lau, Chun Ning
and Bockrath, Marc W.
and Mak, Kin Fai
and Zhang, Fan},
title={Reproducibility in the fabrication and physics of moir{\'e} materials},
journal={Nature},
year={2022},
month={Feb},
day={01},
volume={602},
number={7895},
pages={41-50},
}

@article{
Ribeiro-Palau2018,
author = {Rebeca Ribeiro-Palau  and Changjian Zhang  and Kenji Watanabe  and Takashi Taniguchi  and James Hone  and Cory R. Dean },
title = {Twistable electronics with dynamically rotatable heterostructures},
journal = {Science},
volume = {361},
number = {6403},
pages = {690-693},
year = {2018},
doi = {10.1126/science.aat6981},
}

@article{Yang2019,
    author = {Yang, Yaping and Zou, Yi-Chao and Woods, Colin R. and Shi, Yanmeng and Yin, Jun and Xu, Shuigang and Ozdemir, Servet and Taniguchi, Takashi and Watanabe, Kenji and Geim, Andre K. and Novoselov, Kostya S. and Haigh, Sarah J. and Mishchenko, Artem},
    title = {Stacking Order in Graphite Films Controlled by van
der {W}aals Technology},
    journal = {Nano Lett.},
    volume = {19},
    number = {12},
    pages = {8526-8532},
    year = {2019},
    month = {12},
}

@Article{Hu2022,
author={Hu, Cheng
and Wu, Tongyao
and Huang, Xinyue
and Dong, Yulong
and Chen, Jiajun
and Zhang, Zhichun
and Lyu, Bosai
and Ma, Saiqun
and Watanabe, Kenji
and Taniguchi, Takashi
and Xie, Guibai
and Li, Xiaojun
and Liang, Qi
and Shi, Zhiwen},
title={In-situ twistable bilayer graphene},
journal={Sci. Rep.},
year={2022},
month={Jan},
day={07},
volume={12},
number={1},
pages={204},
}

@article{su1980soliton,
  title={Soliton excitations in polyacetylene},
  author={Su, W-P\_ and Schrieffer, JR and Heeger, AJ},
  journal={Phys. Rev. B},
  volume={22},
  number={4},
  pages={2099},
  year={1980},
  publisher={APS}
}

@article{hagymasi2022observation,
  title={Observation of competing, correlated ground states in the flat band of rhombohedral graphite},
  author={Hagym{\'a}si, Imre and Mohd Isa, Mohammad Syahid and Tajkov, Zolt{\'a}n and M{\'a}rity, Kriszti{\'a}n and Oroszl{\'a}ny, L{\'a}szl{\'o} and Koltai, J{\'a}nos and Alassaf, Assem and Kun, P{\'e}ter and Kandrai, Konr{\'a}d and P{\'a}link{\'a}s, Andr{\'a}s and others},
  journal={Sci. {Adv.}},
  volume={8},
  number={35},
  pages={eabo6879},
  year={2022},
  publisher={American Association for the Advancement of Science}
}

@article{choi2025superconductivity,
  title={Superconductivity and quantized anomalous {Hall} effect in rhombohedral graphene},
  author={Choi, Youngjoon and Choi, Ysun and Valentini, Marco and Patterson, Caitlin L and Holleis, Ludwig FW and Sheekey, Owen I and Stoyanov, Hari and Cheng, Xiang and Taniguchi, Takashi and Watanabe, Kenji and others},
  journal={Nature},
  volume={639},
  number={8054},
  pages={342-347},
  year={2025},
  publisher={Nature Publishing Group UK London}
}

@article{liu2025electronic,
  title={Electronic correlations in rhombohedral graphene at atomic scale},
  author={Liu, Yufeng and Li, Zonglin and Jiang, Shudan and Li, Min and Gu, Yu and Liu, Kai and Shen, Qia and Liu, Liang and Liu, Xiaoxue and Guan, Dandan and others},
  journal={Phys. Rev. Lett.},
  volume={135},
  number={15},
  pages={156401},
  year={2025},
  publisher={APS}
}

@article{zhang2024correlated,
  title={Correlated topological flat bands in rhombohedral graphite},
  author={Zhang, Hongyun and Li, Qian and Scheer, Michael G and Wang, Renqi and Tuo, Chuyi and Zou, Nianlong and Chen, Wanying and Li, Jiaheng and Cai, Xuanxi and Bao, Changhua and others},
  journal={Proc. Natl. Acad. Sci.},
  volume={121},
  number={43},
  pages={e2410714121},
  year={2024},
  publisher={National Academy of Sciences}
}

@article{li2026transdimensional,
  title={Transdimensional anomalous {Hall} effect in rhombohedral thin graphite},
  author={Li, Qingxin and Fan, Hua and Li, Min and Xu, Yinghai and Song, Junwei and Wang, Anqi and Watanabe, Kenji and Taniguchi, Takashi and Chen, Jing-Jing and Tan, Zhenbing and others},
  journal={Nature},
  volume={653},
  pages={384},
  year={2026},
  publisher={Nature Publishing Group UK London}
}

@article{han2025signatures,
  title={Signatures of chiral superconductivity in rhombohedral graphene},
  author={Han, Tonghang and Lu, Zhengguang and Hadjri, Zach and Shi, Lihan and Wu, Zhenghan and Xu, Wei and Yao, Yuxuan and Cotten, Armel A and Sharifi Sedeh, Omid and Weldeyesus, Henok and others},
  journal={Nature},
  volume={643},
  pages={654-661},
  year={2025},
  publisher={Nature Publishing Group UK London}
}

@article{soneji2026rhombohedral,
  title={Rhombohedral graphite junctions as a platform for continuous tuning between topologically trivial and non-trivial electronic phases},
  author={Soneji, Luke and Crampin, Simon and Mucha-Kruczy{\'n}ski, Marcin},
  journal={Commun. Phys.},
  volume={9},
  pages={190},
  year={2026},
  publisher={Nature Publishing Group UK London}
}

@article{soneji2026electron,
  title={Electron localization, charge redistribution, and emergence of topological states at graphite junctions},
  author={Soneji, Luke and Crampin, Simon and Mucha-Kruczynski, Marcin},
  journal={arXiv preprint arXiv:2604.11515},
  year={2026}
}

@article{soler2002siesta,
  title={The {SIESTA} method for ab initio order-N materials simulation},
  author={Soler, Jos{\'e} M and Artacho, Emilio and Gale, Julian D and Garc{\'\i}a, Alberto and Junquera, Javier and Ordej{\'o}n, Pablo and S{\'a}nchez-Portal, Daniel},
  journal={{J. Phys.: Condens. Matter}},
  volume={14},
  number={11},
  pages={2745},
  year={2002}
}

@article{garcia2020siesta,
  title={{SIESTA}: Recent developments and applications},
  author={Garc{\'\i}a, Alberto and Papior, Nick and Akhtar, Arsalan and Artacho, Emilio and Blum, Volker and Bosoni, Emanuele and Brandimarte, Pedro and Brandbyge, Mads and Cerd{\'a}, Jorge I and Corsetti, Fabiano and others},
  journal={J. Chem. Phys.},
  volume={152},
  number={20},
  pages={204108},
  year={2020},
  publisher={AIP Publishing}
}

@article{garcia2018psml,
  title={The psml format and library for norm-conserving pseudopotential data curation and interoperability},
  author={Garc{\'\i}a, Alberto and Verstraete, Matthieu J and Pouillon, Yann and Junquera, Javier},
  journal={Comput. Phys. Commun.},
  volume={227},
  pages={51-71},
  year={2018},
  publisher={Elsevier}
}

@article{brandbyge2002density,
  title={Density-functional method for nonequilibrium electron transport},
  author={Brandbyge, Mads and Mozos, Jos{\'e}-Luis and Ordej{\'o}n, Pablo and Taylor, Jeremy and Stokbro, Kurt},
  journal={Phys. Rev. B},
  volume={65},
  number={16},
  pages={165401},
  year={2002},
  publisher={APS}
}

@article{papior2017improvements,
  title={Improvements on non-equilibrium and transport {Green} function techniques: The next-generation transiesta},
  author={Papior, Nick and Lorente, Nicol{\'a}s and Frederiksen, Thomas and Garc{\'\i}a, Alberto and Brandbyge, Mads},
  journal={Comput. Phys. Commun.},
  volume={212},
  pages={8-24},
  year={2017},
  publisher={Elsevier}
}

@article{lee2010higher,
  title={Higher-accuracy van der {Waals} density functional},
  author={Lee, Kyuho and Murray, {\'E}amonn D and Kong, Lingzhu and Lundqvist, Bengt I and Langreth, David C},
  journal={Phys. Rev. B},
  volume={82},
  number={8},
  pages={081101},
  year={2010},
  publisher={APS}
}

@article{koshino2009trigonal,
  title={Trigonal warping and {Berry}’s phase {N}$\pi$ in {ABC}-stacked multilayer graphene},
  author={Koshino, Mikito and McCann, Edward},
  journal={Phys. Rev. B},
  volume={80},
  number={16},
  pages={165409},
  year={2009},
  publisher={APS}
}

@book{asboth2016short,
  title={A short course on topological insulators},
  author={Asb{\'o}th, J{\'a}nos K and Oroszl{\'a}ny, L{\'a}szl{\'o} and P{\'a}lyi, Andr{\'a}s},
  volume={919},
  year={2016},
  publisher={Springer}
}

@article{mccann2006landau,
  title={Landau-level degeneracy and quantum {Hall} effect in a graphite bilayer},
  author={McCann, Edward and Fal’ko, Vladimir I},
  journal={{Phys. Rev. Lett.}},
  volume={96},
  number={8},
  pages={086805},
  year={2006},
  publisher={APS}
}

@article{taut2014electronic,
  title={Electronic structure of stacking faults in rhombohedral graphite},
  author={Taut, M and Koepernik, K and Richter, M},
  journal={Phys. Rev. B},
  volume={90},
  number={8},
  pages={085312},
  year={2014},
  publisher={APS}
}

@article{taut2013electronic,
  title={Electronic structure of stacking faults in hexagonal graphite},
  author={Taut, M and Koepernik, K and Richter, M},
  journal={Phys. Rev. B},
  volume={88},
  number={20},
  pages={205411},
  year={2013},
  publisher={APS}
}

@article{weht2026topological,
  title={Topological flat bands emerging at the inversion of stacking order in rhombohedral graphite},
  author={Weht, Ruben and Aligia, Armando A and Nunez-Regueiro, Manuel},
  journal={arXiv preprint arXiv:2605.01115},
  year={2026}
}

@article{kaladzhyan2021surface,
  title={Surface states and quasiparticle interference in {Bernal} and rhombohedral graphite with and without trigonal warping},
  author={Kaladzhyan, Vardan and Pinon, Sarah and Joucken, Fr{\'e}d{\'e}ric and Ge, Zhehao and Quezada-Lopez, Eberth A and Taniguchi, T and Watanabe, K and Velasco Jr, Jairo and Bena, Cristina},
  journal={Phys. Rev. B},
  volume={104},
  number={15},
  pages={155418},
  year={2021},
  publisher={APS}
}

@article{troullier1991efficient,
  title={Efficient pseudopotentials for plane-wave calculations},
  author={Troullier, Norman and Martins, Jos{\'e} Lu{\'\i}s},
  journal={Phys. Rev. B},
  volume={43},
  number={3},
  pages={1993},
  year={1991},
  publisher={APS}
}

@misc{zerothi_sisl,
  author       = {Papior, Nick and Febrer, Pol},
  title        = {sisl: v 0.16.2},
  year         = {2026},
  doi          = {10.5281/zenodo.597181},
  url          = {https://doi.org/10.5281/zenodo.597181},
}

@article{birowska2011van,
  title={Van der {Waals} density functionals for graphene layers and graphite},
  author={Birowska, Magdalena and Milowska, Karolina and Majewski, J},
  journal={Acta Physica Polonica A},
  volume={120},
  number={5},
  pages={845-848},
  year={2011},
  publisher={Polska Akademia Nauk. Instytut Fizyki PAN}
}

@article{lee2008growth,
  title={The growth of {AA} graphite on (111) diamond},
  author={Lee, Jae-Kap and Lee, Seung-Cheol and Ahn, Jae-Pyoung and Kim, Soo-Chul and Wilson, John IB and John, Phillip},
  journal={J. Chem. Phys.},
  volume={129},
  number={23},
  pages={234709},
  year={2008},
  publisher={AIP Publishing}
}

@article{francas2025influence,
  title={Influence of screw dislocations on stacking order in graphite},
  author={Francas, Gabriel R and Delenda, Anouar A and Martin, Jacob W and Bousige, Colin and Suarez-Martinez, Irene and Marks, Nigel A and Ewels, Chris},
  journal={Carbon},
  volume={247},
  pages={120995},
  year={2026},
  publisher={Elsevier}
}

@article{PhysRevB.73.245426,
  title = {Electronic states and {Landau} levels in graphene stacks},
  author = {Guinea, F. and Castro Neto, A. H. and Peres, N. M. R.},
  journal = {Phys. Rev. B},
  volume = {73},
  issue = {24},
  pages = {245426},
  numpages = {8},
  year = {2006},
  month = {Jun},
  publisher = {American Physical Society},
  doi = {10.1103/PhysRevB.73.245426},
  url = {https://link.aps.org/doi/10.1103/PhysRevB.73.245426}
}

@article{shi2020electronic,
  title={Electronic phase separation in multilayer rhombohedral graphite},
  author={Shi, Yanmeng and Xu, Shuigang and Yang, Yaping and Slizovskiy, Sergey and Morozov, Sergey V and Son, Seok-Kyun and Ozdemir, Servet and Mullan, Ciaran and Barrier, Julien and Yin, Jun and others},
  journal={Nature},
  volume={584},
  number={7820},
  pages={210-214},
  year={2020},
  publisher={Nature Publishing Group UK, London}
}

@article{slizovskiy2019films,
  title={Films of rhombohedral graphite as two-dimensional topological semimetals},
  author={Slizovskiy, Sergey and McCann, Edward and Koshino, Mikito and Fal’ko, Vladimir I},
  journal={Commun. Phys.},
  volume={2},
  number={1},
  pages={164},
  year={2019},
  publisher={Nature Publishing Group UK, London}
}

@article{henck2018flat,
  title={Flat electronic bands in long sequences of rhombohedral-stacked graphene},
  author={Henck, Hugo and Avila, Jose and Ben Aziza, Zeineb and Pierucci, Debora and Baima, Jacopo and Pamuk, Bet{\"u}l and Chaste, Julien and Utt, Daniel and Bartos, Miroslav and Nogajewski, Karol and others},
  journal={Phys. Rev. B},
  volume={97},
  number={24},
  pages={245421},
  year={2018},
  publisher={APS}
}

@article{PhysRevB.89.035431,
  title = {Ab initio quasiparticle band structure of {ABA} and {ABC}-stacked graphene trilayers},
  author = {Menezes, Marcos G. and Capaz, Rodrigo B. and Louie, Steven G.},
  journal = {Phys. Rev. B},
  volume = {89},
  issue = {3},
  pages = {035431},
  numpages = {5},
  year = {2014},
  month = {Jan},
  publisher = {American Physical Society},
  doi = {10.1103/PhysRevB.89.035431},
  url = {https://link.aps.org/doi/10.1103/PhysRevB.89.035431}
}

@article{sarsfield2025electronic,
  title={Electronic properties of stacking faults in {Bernal} graphite},
  author={Sarsfield, Patrick Johansen and Slizovskiy, Sergey and Koshino, Mikito and Fal’ko, Vladimir},
  journal={npj Comput. Mater.},
  volume={11},
  number={1},
  pages={142},
  year={2025},
  publisher={Nature Publishing Group UK, London}
}

@article{pizzochero2021edge,
  title={Edge disorder in bottom-up zigzag graphene nanoribbons: implications for magnetism and quantum electronic transport},
  author={Pizzochero, Michele and Barin, Gabriela Borin and Cernevics, Kristians and Wang, Shiyong and Ruffieux, Pascal and Fasel, Roman and Yazyev, Oleg V},
  journal={{J. Phys. Chem. Lett.}},
  volume={12},
  number={19},
  pages={4692-4696},
  year={2021},
  publisher={ACS Publications}
}

@article{pizzochero2021quantum,
  title={Quantum electronic transport across ‘bite’ defects in graphene nanoribbons},
  author={Pizzochero, Michele and {\v{C}}er{\c{n}}evi{\v{c}}s, Kristi{\=a}ns and Borin Barin, Gabriela and Wang, Shiyong and Ruffieux, Pascal and Fasel, Roman and Yazyev, Oleg V},
  journal={2D Mater.},
  volume={8},
  number={3},
  pages={035025},
  year={2021},
  publisher={IOP Publishing}
}

@article{han2024large,
  title={Large quantum anomalous {Hall} effect in spin-orbit proximitized rhombohedral graphene},
  author={Han, Tonghang and Lu, Zhengguang and Yao, Yuxuan and Yang, Jixiang and Seo, Junseok and Yoon, Chiho and Watanabe, Kenji and Taniguchi, Takashi and Fu, Liang and Zhang, Fan and others},
  journal={Science},
  volume={384},
  number={6696},
  pages={647-651},
  year={2024},
  publisher={American Association for the Advancement of Science}
}

@article{parra2026nematic,
  title={Nematic and partially polarized phases in rhombohedral graphene with varying number of layers: An extensive {Hartree-Fock} study},
  author={Parra-Mart{\'\i}nez, Guillermo and Jimeno-Pozo, Alejandro and Silva-Guill{\'e}n, Jose Angel and Guinea, Francisco},
  journal={Adv. Mater.},
  volume={38},
  number={51},
  pages={e14008},
  year={2026},
  publisher={Wiley Online Library}
}

@article{miao2026various,
  title={Various electronic crystal phases in rhombohedral graphene multilayers},
  author={Miao, Wangqian and Li, Chu},
  journal={Phys. Rev. B},
  volume={113},
  number={15},
  pages={155136},
  year={2026},
  publisher={APS}
}

@article{zhou2024layer,
  title={Layer-polarized ferromagnetism in rhombohedral multilayer graphene},
  author={Zhou, Wenqiang and Ding, Jing and Hua, Jiannan and Zhang, Le and Watanabe, Kenji and Taniguchi, Takashi and Zhu, Wei and Xu, Shuigang},
  journal={Nat. Commun.},
  volume={15},
  number={1},
  pages={2597},
  year={2024},
  publisher={Nature Publishing Group UK, London}
}
\end{document}